\documentclass[twocolumn]{revtex4}
\usepackage{graphicx}

\begin{document}

\title{Spin Structure and Critical Nucleation Frequency of
Fractionalized Vortices
in 2D Topologically Ordered Superfluids of Cold Atoms}
%

%

\author{Jun Liang Song and Fei Zhou}
\affiliation{Department of Physics and Astronomy,
The University of British Columbia, Vancouver, B. C., Canada V6T1Z1}
\date{{\small \today}}

%

\begin{abstract}
We have studied spin structures of fluctuation-driven fractionalized vortices
and topological spin order in
2D nematic superfluids of cold sodium atoms.
Our Monte Carlo simulations suggest
a {\em softened} $\pi$-spin disclination structure
in a half-quantum vortex when spin correlations are short
ranged; in addition, calculations indicate
that a unique {\em non-local} topological spin order
emerges simultaneously as cold atoms become a superfluid below a critical
temperature.
We have also estimated fluctuation-dependent
critical frequencies for half-quantum vortex nucleation in rotating optical traps and discussed
probing these excitations in experiments.
\end{abstract}
\maketitle

Quantum number fractionalization has been one of the most
fundamental and exciting concepts studied in modern many-body
physics and topological field theories\cite{Su80,Tsui82,Laughlin83,Jackiw76}.
During the past few years, low dimensional fractionalized quantum states have
further been proposed to be promising candidates for carrying out fault
tolerant quantum computation\cite{Kitaev03} and their realizations in optical lattices were
explored\cite{Duan03,Buchler05,Micheli06}.
A closely related topic in which
there has also been a growing interest
is vortex fractionalization in
cold gases (see for instance Ref.\cite{Demler02,Mukerjee06,Semenoff07,Thouless98}).
Especially, motivated by experiments on low dimensional
cold gases\cite{Hadzibabic06},
{\em Mukerjee et
al} studied 2D superfluids of cold atoms and
analyzed the role played by fractionalized vortices
in phase transitions\cite{Mukerjee06}.
However, spin structures of those half-quantum vortices induced by thermal fluctuations
and potential topological order\cite{Wen04}
haven't been thoroughly explored and remain to be understood.
In this Letter
we illustrate spin structures of
fractionalized vortices; in addition, we also show that 2D quantum gases
with short ranged spin correlations can have a topological spin order.
We further estimate critical nucleation frequencies of fractionalized
vortices in optical traps
which can potentially be studied in experiments\cite{Stenger98}.

Our simulations illustrate
that in a fundamental vortex carrying
one-half of circulation
quantum $h/m$ ($h$ is the Planck constant and $m$ is the mass of atoms),
there exists a {\em softened} spin disclination
(i.e. a disclination in the absence of spin stiffness) even when
local spin moments are strongly fluctuating at finite temperatures.
The topological winding
number associated with softened spin disclinations is
conserved as far as the phase rigidity remains finite.
This effectively leads to a {\em non-local} topological spin order.
Such a nonlocal order is absent in a conventional condensate of atoms or
pairs of atoms.
We have further studied
creation of these excitations in rotating
superfluids and obtained fluctuation-dependent
critical frequencies for half-quantum vortex({\em HQV}) nucleation.
{\em HQV}s in
traps can be probed
by measuring a
precession of eigenaxes of surface quadrupole modes.

We employ the
Hamiltonian introduced previously for F=1 sodium atoms in
optical lattices\cite{Zhou03,Demler02},
\begin{eqnarray}
{\cal H}&=& \sum_{k} \frac{b_L}{2} \hat{\rho}^2_k +\frac{c_L}{2}
\hat{\cal S}^2_k
\nonumber \\
&-& t_L\sum_{<kl>} (\psi^\dagger_{k,\alpha}
\psi^{~}_{l,\alpha} + h.c.) - \sum_k \mu \hat{\rho}_k.
\label{Hamiltonian}
\end{eqnarray}
Here $k$ is the lattice site index and $<kl>$ are the nearest
neighbor sites. $\mu$ is the chemical potential and $t_L$ is the
one-particle hopping amplitude. Two coupling constants
are $b_L (c_L) =b(c)\frac {4\pi\hbar^2}{m} \int d{\bf r}
(\phi_w^*({\bf r})\phi_w({\bf r}))^2$; $b$,
$c$ are effective s-wave scattering lengths, $\phi_w$ is
the localized Wannier function for atoms in a periodical potential.
Operators
$\psi^\dagger_\alpha$, $\alpha=x,y,z$ create hyperfine spin-one
atoms in $\frac{1}{\sqrt{2}}(|1>-|-1>)$, $-i \frac{1}{\sqrt{2}}({|1>+|-1>})$ and
$|0>$ states respectively. The spin and number operators are
defined as $\hat{S}_\alpha=-i\epsilon_{\alpha\beta\gamma}\psi^\dagger_\beta\psi_\gamma$,
and $\hat{\rho}=\psi^\dagger_\alpha\psi_\alpha$.
Spin correlations are mainly induced by
interaction $c_L \hat{\cal S}^2$.
Here we consider antiferromagnetic spin-dependent interactions such as in sodium atoms where $c_L > 0$.
Minimization of this antiferromagnetic spin-dependent interaction
requires that
the order parameter $\Psi_\alpha(=<\psi^\dagger_{k,\alpha}>)$
be a {\em real} vector up
to a global phase, i.e. $\Psi=\sqrt{N}{\bf n}\exp(i\chi)$ where ${\bf n}$ is
a unit director on a two-sphere, $exp(i\chi)$ represents a phase
director and $N$ is the number of atoms per site.
All low energy degrees of freedom are characterized by
configurations where ${\bf n}$ and $\chi$ vary slowly in space and
time\cite{Demler02}. Low lying collective
modes include spin-wave
excitations with energy dispersion $\omega(q)=v_s q$,
$v_s=\sqrt{c_L t_L}a$ and
phase-wave excitations with $\omega(q)=v_p q$,
$v_p=\sqrt{b_L t_L} a$ (here $a$ is the lattice constant).
In one-dimensions, low energy
{\em quantum} fluctuations destroy spin order
leading to quantum spin disordered
superfluids\cite{Zhou01}.
In two-dimensions, the amplitude of quantum spin fluctuations
is of order of $c_L/t_L$ and is negligible in
shallow lattices as $t_L$ is order-of-magnitude bigger than
$c_L$.
At finite temperatures,
spin correlations are mainly driven by {\em long wave length}
thermal fluctuations,
analogous to quantum $1D$ cases.
This aspect was also paid attention to
previously and normal-superfluid transitions were
investigated\cite{Mukerjee06}.

\begin{figure}[t]
\includegraphics[width=\columnwidth]{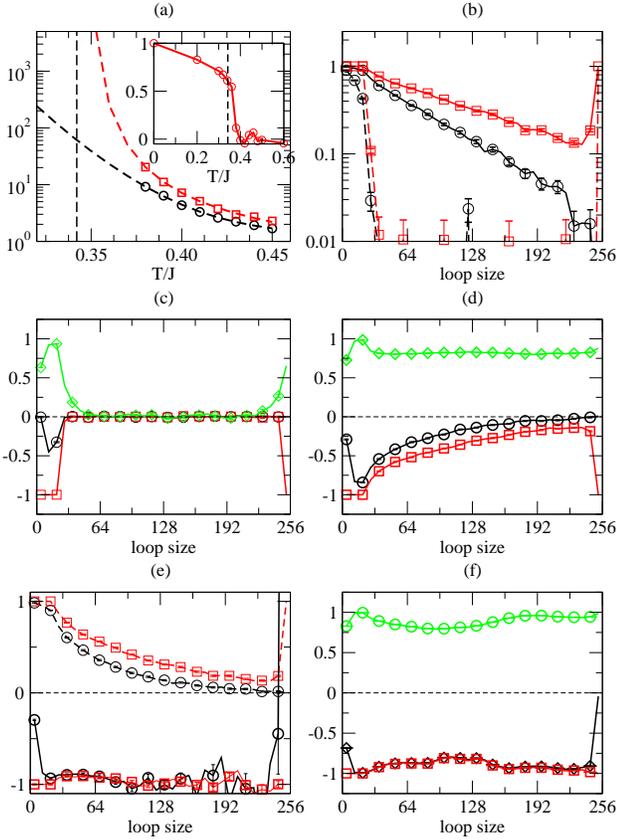}
\caption{ (color online)
a)Spin(circle) and phase(square) correlation lengths versus
temperatures; inset is for the renormalized phase coupling constant
$J_p$(in units of $J$).
Dashed lines are the fits to $\exp(A/\sqrt{T-T_c})$ for phase correlations
with $T_c\approx 0.35J$ and $\exp(B/T)$ for spin correlations.
b) Thermal average $\langle W_{s} \rangle$(circle) and $\langle W_p \rangle$(square) versus the loop
perimeter at $T=0.33J$(solid line) and
$T=0.45J$(dashed line).
c)-f): $\langle W_{s} \rangle_{hv}$(circle), $\langle W_p \rangle_{hv}$(square) and $\langle C
\rangle_{hv}$(diamond)
averaged over
configurations with {\em HQV} boundary conditions.
c)-d),f) are for $T/J=0.45$,$0.33$,$0.20$ respectively;
in e), we also show
$\langle W_{s,p}\rangle_{hv}$ normalized in terms of back ground values $\langle W_{s,p}\rangle_{bg}$ at
$T=0.33J$.
$\langle W_{s,p}\rangle_{bg}$(dashed lines) are averaged over configurations with
uniform phases at the boundary.}
\label{fig1}
\end{figure}

We therefore study the following Hamiltonian which effectively captures
long wave length thermal fluctuations

\begin{eqnarray}
H=-\sum_{<kl>} J_{kl} {\bf n}_k \cdot {\bf n}_l {\bf \Phi}_k\cdot
{\bf \Phi}_l; \label{effHamiltonian}
\end{eqnarray}
here states at each site are specified by two unit directors: a nematic
director,
${\bf n}=(\cos\phi_0\sin\theta_0, \sin\phi_0\sin\theta_0,\cos\theta_0)$ and a
phase director, ${\bf \Phi}=(\cos\chi,\sin\chi)$. $J_{kl}=2{N}t_L$ is
the effective coupling between two neighboring sites and depends on $N$,
the number of atoms per site.
The model is invariant under the following
{\em local} Ising gauge
transformation: ${\bf n}_i \rightarrow  s_i {\bf n}_i$, ${\bf \Phi}
\rightarrow s_i {\bf \Phi}_i$, and $s_i=\pm 1$.
In the following, we present results of our
simulations on 2D superfluids, especially spin
structures, energetics of {\em HQV}s
and nucleation of {\em HQV}s in rotating optical traps
using the effective Hamiltonian in Eq.\ref{effHamiltonian}.

{\em HQV and underlying topological spin order}
Around a {\em HQV}, both phase and nematic directors
rotate slowly by $180^0$;
in polar coordinates $(\theta, \rho)$, a {\em HQV} in condensates is
represented by $\Psi(\theta,
\rho)=\sqrt{N(\rho)} \exp(i\theta/2) {\bf n}(\theta)$, with ${\bf
n}=(\cos(\theta/2),\sin(\theta/2),0)$\cite{Thouless98,Zhou01}. The
question
here is
whether,
when nematic directors are not ordered, a spin disclination
is still present in a {\em HQV}.
To fully take into account $2D$ thermal fluctuations, we carry out Monte Carlo simulations on a
square lattice of $128 \times 128$ sites
and study spatial correlations between a {\em HQV}
and a $\pi$-spin disclination, and topological order.

We first identify critical temperatures of the normal-superfluid
phase transition by calculating correlations and the phase rigidity.
The gauge-invariant quadrupole-quadrupole correlation functions
we have studied are

\begin{equation}
f^{s,p}({\bf r}_1, {\bf r}_2)=<Q^{s,p}_{\alpha\beta}({\bf r}_1)
Q^{s,p}_{\alpha\beta}({\bf r}_2)>.
\end{equation}
Here $Q^{s}_{\alpha\beta}({\bf r}_1)={\bf n}_{1,\alpha}{\bf n}_{1,\beta}
-(1/3) \delta_{\alpha\beta}$, $\alpha=x,y,z$; $Q^{p}_{\alpha\beta}({\bf r}_1)
={\bf \Phi}_{1,\alpha}{\bf \Phi}_{1,\beta}-(1/2)\delta_{\alpha\beta}$, $\alpha=x,y$.
In simulations, we have studied these correlation functions and found
that the phase correlation length for $f^{p}({\bf r}_1,{\bf r}_2)$
becomes divergent at a temperature $0.35J$ which is identified as a critical temperature
$T_c$. We also calculate the phase rigidity or the renormalized phase coupling $J_p$

\begin{eqnarray}
J_p=\frac{\partial^2  F}{\partial \delta \chi^2};
\end{eqnarray}
here $\delta \chi$ is a small phase difference applied
across the opposite boundaries of the lattice and $F$ is the corresponding free energy.
We indeed find that it approaches zero at $T_c$ while at $T=0$ $J_p$ takes a bare value $J$.
Meanwhile, the spin correlation function
$f^s({\bf r}_1,{\bf r}_2)$ remains to be short ranged across $T_c$. By extrapolating
our data to lower temperatures, we find that the spin correlation length
diverges only at $T=0$ (see Fig.\ref{fig1}a).
Our simulations for correlation lengths are in agreement with
previous results in Ref.\cite{Mukerjee06};
they are also consistent with
the continuum limit of the
model in Eq.\ref{effHamiltonian}
which is equivalent to an $XY$ model and an $O(3)$
nonlinear-sigma model.

In order to keep track of the winding of nematic directors
in a wildly fluctuating back ground,
we introduce the following gauge invariant $\pi$-rotation
checking operator, which is essentially a product of
sign-checking operators

\begin{eqnarray}
W_s=\prod_{<kl>\in \cal C} sign({\bf n}_k \cdot {\bf n}_l),
W_p=\prod_{<kl>\in \cal C} sign({\bf \Phi}_k \cdot {\bf \Phi}_l).
\end{eqnarray}
Here the product is carried out
along a closed square-shape path ${\cal C}$ centered at the origin of a $2D$ lattice.
$W_{s,p}$ can be either $+1$ or $-1$; and $W_{s}$($W_p$) is $-1$
when ${\cal C}$ encloses a $\pi$-spin disclination
({\em HQV}). The gauge invariant circulation of supercurrent velocity (in units of $\pi\hbar/m$) is defined as
$C=\frac{1}{\pi}\sum_{\in {\cal C}}
sign ({\bf n}_k\cdot {\bf n}_l)\sin(\chi_k-\chi_l)$;
this quantity is equal to one in a {\em HQV}.

In our simulations,
we investigate the winding number $<W_{s,p}>_{hv}$
averaged over configurations where
phase directors rotate by $180^0$ around the boundary of the
lattice and the center plaquette.
At temperatures above the normal-superfluid transition temperature $T_c$,
both winding numbers $W_{s,p}$ and circulation $C$ are averaged to zero within our
numerical accuracy(see Fig.\ref{fig1}).
And our choice of boundary conditions does not lead
to a vortex or disclination configuration in the absence of
phase rigidity.
Below $T_c$, the circulation $C$ is averaged to one indicating that the
boundary conditions effectively project out {\em HQV}
configurations. Meanwhile,
we observe loop-perimeter dependent $\langle W_{s,p} \rangle_{hv}$ which
can be attributed to the background fluctuations of
{\em HQV} or disclination pairs. The loop-perimeter dependence of $\langle W_{s,p}\rangle_{hv}$ here is almost identical to that for
uniform boundary conditions, i.e. the back ground value.
After normalizing $\langle W_{s,p}\rangle_{hv}$ in terms of
background winding numbers $\langle W_{s,p} \rangle_{bg}$, we find both $\frac{\langle W_{s}\rangle_{hv}}{\langle
W_{s}\rangle_{bg}}$
and
$\frac{\langle W_p \rangle_{hv}}{\langle W_{p}\rangle_{bg}}$ approach ${-1}$ (see Fig.\ref{fig1}).
We thus demonstrate that a softened disclination
is spatially correlated with a {\em HQV}.
At the temperatures we carry out these simulations
the spin correlation length is sufficiently short compared to the size of
the lattice.
At further lower temperatures, the spin correlation length becomes
longer than the lattice size and fluctuations of
pairs of disclination-anti-disclination are strongly suppressed;
$<W_{s,p}>_{hv}$ are equal to $-1$ for almost all loops, which corresponds
to a mean field result.

Results in Fig.\ref{fig1}
indicate that there exists a softened spin disclination in a {\em HQV}.
This is a distinct feature in our systems and there exist no
such additional magnetic
structures in {\em HQV}s in conventional molecular condensates of atom
pairs discussed perviously\cite{Romans04}.
Thus, $\pi$-disclinations like {\em HQV}s have
logarithmically divergent
energies and are fully suppressed in ground states.
Our results also illustrate that although the average local
spin quadrupole moments $Q^s_{\alpha\beta}$ vanish because of strong
fluctuations,
an overall $\pi$-rotation of nematic directors in disclinations is still conserved
because of a coupling to the superfluid component.
This coupling between a {\em HQV} and disclination can also be
attributed\cite{Song08} to a coupling between Higgs matter and discrete
gauge fields\cite{Fradkin79}. Furthermore,
the absence of unbound $\pi$-disclinations in superfluids indicates
a topological order, similar to the one introduced previously
for an isotropic phase of liquid crystal\cite{Toner93}.
Consequently, once a conventional
phase order appears below a critical temperature,
a topological spin order simultaneously emerges
while spin correlations remain short ranged.

The emergent topological order can be further verified by
examining the average of product-operator
$W_{s,p}$ over all configurations (with open boundaries).
Above the normal-superfluid
transition temperature $T_c$,
we again find that $W_{s,p}$ both are averaged to zero within
our numerical accuracy implying proliferation of unbound
{\em HQV}s or disclinations.
Below $T_c$, we study the loop-perimeter dependence of average winding numbers $\langle W_{s,p} \rangle$ and find that
both $\ln \langle W_p \rangle$ and $\ln \langle W_s \rangle$ are linear functions of loop-perimeter
analogous to the Wilson-loop-product of deconfining gauge fields\cite{Wilson74};
if there were unbound disclinations, one should expect that
$\ln \langle W_p \rangle$ is
proportional to, instead of the loop-perimeter,
the loop-area which represents the number of unbound disclinations enclosed by the loop.

{\em Critical frequency for HQV nucleation}
Let us now turn to the nucleation of those excitations in rotating
traps\cite{Madison00,Haljan01,AboShaeer01,Fetter01,Dalfovo01,
Tsubota02,Isoshima02}.
To understand the critical frequency for
nucleation, we study
the free energy of a vortex, in a rotating frame, as a function of the
distance $r$ from the axis of a cylindrical optical trap (the axis is
along the $z$-direction),

\begin{eqnarray}
F_{h.v.}(r)=F^0_{h.v.}(r)-\Omega L_z(r).
\end{eqnarray}
Here $F^0_{h.v.}(r)$ is the free energy of a {\em HQV} located at distance
$r$ from the trap axis in the absence of rotation,
$L_z(r)$ is the angular
momentum of the vortex state and $\Omega$ is the rotating frequency.

In a $2D$ lattice without a trapping potential,
$F^0_{h.v.}$ is approximately equal to $\frac{\pi}{4}(J_p + J_s)
\ln (L/a)$, with leading contributions
from phase winding and spin twisting;
here $J_{p,s}$ are renormalized phase and spin
coupling respectively and $L$ is the size of system.
For an integer-quantum vortex ({\em IQV}),
$F^0_{v}$ is equal to $\pi J_p \ln (L/a)$.
The ratio between $F^0_{h.v.}$ and $F^0_{v}$
depends on the ratio $J_s/J_p$ or spin
fluctuations;
in the limit $L$ approaches infinity,
the ratio $F^0_{h.v.}/F^0_v$ changes discontinuously
from $\frac{1}{2}$
at $T=0$ where $J_p\approx J_s =J$ to
$\frac{1}{4}$ at finite low temperatures in $2D$ where $J_s$ vanishes.
In simulations of a finite trap (see below),
because of a finite size effect we find that this ratio varies from
$0.5$ to $0.2$ smoothly as temperatures
increase from $0$ to $T_c$.

\begin{figure}[t]
\includegraphics[width=\columnwidth]{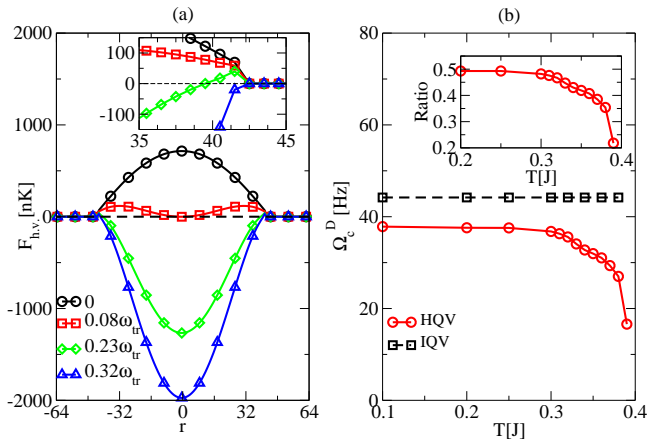}
\caption{ (color online)
a) The free energy of a {\em HQV}
at a distance $r$ from the trap center at rotation
frequencies $0$, $0.08$, $0.23$ and $0.32$ (in units of trap frequency $\omega_{tr}=120Hz$)
at $T=0.33J$. Details near the edge are shown in the inset.
At the center, the exchange coupling is $J=154$ nK.
b) The critical frequencies $\Omega^D_c$ for {\em HQV}s (solid line)
and {\em IQV}s (dashed line).
Inset is the temperature-dependence of the ratio between the {\em HQV}
energy $F^0_{h.v.}$
and the {\em IQV} energy $F^0_{v}$.}
\label{fig3}
\end{figure}

To study nucleation of half-quantum vortices in an optical trap,
we assume a nearly harmonic trapping potential
$V(r)=1/2 m \omega^2_{tr} r^2$, with $\omega_{tr}$ being the trap frequency.
The average number of particles per site $N(r)$ has
a Thomas-Fermi profile; $N(r)=N_0(1-r^2/R_{TF}^2)$, here $N_0$ is the number density at the center and
$R_{TF}$ is the Thomas-Fermi radius.
Furthermore,
the optical lattice potential
along the axial direction is sufficiently deep so that atoms are confined
in a two-dimensional $xy$ plane;
the in-plane lattice potential depth is set to be
$5E_R$($E_R$ is the photon recoil energy).

For the trap and lattice geometry described above, we calculate parameters
in Eq.\ref{Hamiltonian} and obtain
$t_L=77$nK,
$b_L=187$nK and
$c_L=10$nK.
For $N_0=1.2$ and trap frequency $\omega_{tr}=120Hz$,
we also find that $R_{TF}=43a$ and
$L_{h.o.}=4a$ where
$L_{h.o.}(=1/\sqrt{2 m \omega_{tr}})$ is the harmonic oscillator
length.
The coupling $J_{kl}$ in Eq.\ref{effHamiltonian} depends on the distance
from the center of
trap and at the center, the coupling is about $154$nK.
In non-rotating or slowly rotating traps,
the free energy maximum is located at the center
and there should be no vortices in the trap.
As frequencies are increased, a local energy minimum
appears at the center and becomes degenerate with the no-vortex state
at a thermodynamic critical frequency (which is about $0.08\omega_{tr}$ at $T=0.33J$);
however because of a large energy barrier separating the two degenerate states as shown in Fig.\ref{fig3},
vortices are still prohibited from entering the trap.

Further speeding up rotations results in an energetically lower and spatially
narrower barrier.
Within the range of temperatures studied,
thermal activation turns out to be insignificant within an experimental time scale ($\sim
100ms$)
because of low attempt frequencies.
So only when the spatial
width of barrier becomes comparable to a hydrodynamic
breakdown length\cite{Feder00},
the barrier can no longer be felted and
vortices start to penetrate into the trap.
We use this criterion to numerically determine
the dynamical critical frequency for vortex nucleation $\Omega^D_c$;
for {\em IQV}s, the calculated $\Omega^D_c$ is a flat function
of $T$ (see Fig.\ref{fig3}b) which is qualitatively consistent with earlier estimates\cite{Simula02}.

For {\em HQV}s, $F^0_{h.v.}$ depends on the renormalized spin coupling $J_s$
and therefore the amplitude of spin
fluctuations. Because of this,
$\Omega^D_c$  varies from about $0.32 \omega_{tr}$
at $T=0$ where $J_s \approx J_p$ and $F_{h.v.}\approx 0.5 F_{v}$ due to
a finite size effect (see the inset of Fig.\ref{fig3}), to
about $0.17 \omega_{tr}$
at temperatures close to $T_c$ where $J_s=0$ and $F_{h.v.}\approx 0.2 F_{v}$.
In other words, this unique temperature dependence can be considered to
be an indicator of fluctuation-driven vortex fractionalization.
It is worth remarking that in the thermodynamic limit
where $J_s$ approaches zero at any finite temperatures,
$F_{h.v.}$ approaches $\frac{1}{4} F_{v}$ as mentioned before.
Consequently, $\Omega^D_c$(about $0.17 \omega_{tr}$) for {\em HQV}s is about
one-half of the critical frequency for IQVs (about $0.36\omega_{tr}$ for the finite trap
studied here). Also note that the zero temperature estimate of $\Omega^D_c$ is
close to the previously obtained value of critical
frequencies of {\em HQV}s in Bose-Einstein
condensates\cite{Isoshima02}.

The interaction between two {\em HQV}s with the same
vorticity at a separation distance $d$ contains two parts.
One, $V_{cc} (>0)$ is from interactions between
two supercurrent velocity fields which is
logarithmic as a function of $d$;
and the other, $V_{ss}$ is from interaction between two spin twisting
fields
accompanying {\em HQV}s.
For a disclination-anti disclination pair,
in the dilute limit
one finds that
$V_{cc}\sim - V_{ss}$ resulting in a cancellation of long range interactions.
The resultant short-range repulsions lead to square vortex lattices found in numerical
simulations\cite{Andrew08}.
For fluctuation-driven fractionalized vortices,
$V_{ss}$ is almost zero and the overall interactions are
always logarithmically repulsive.
{\em HQV}s nucleated
in a rotating trap should therefore form a
usual triangular vortex lattice.

Individual vortex lines can be probed either by studying a precession of eigenaxes of
surface quadrupole mode
in rotating superfluids\cite{Chevy00}.
In the later approach, one studies the angular momentum carried per
particle in a {\em HQV} state. When a {\em HQV} is nucleated in the
trap,
superfluids are no longer irrotational and
the angular momentum per particle is $h/2$
rather than $h$ per particle for an integer vortex state.
When a surface quadrupole oscillation across a rotating superfluid is
excited,
larger axes of quadrupole oscillation start
to precess just as in the case of
integer vortices. However, the precession rate is only one half of the value for
an integer vortex state which can be studied in experiments.

In conclusion, 2D superfluids of sodium atoms have a non-local
topological spin order. In rotating traps,
fluctuation-driven fractionalized vortices
can nucleate at a critical frequency which is about half of that for integer vortices.
Observation of these exotic excitations could substantially improve our
understanding of topological order and
fractionalization.
We thank J. Zhang and Z. C. Gu for
contributions at an early stage of the project.
This work is in part supported
by the office of the Dean of Science, UBC, NSERC (Canada), Canadian
Institute for Advanced Research, and the A. P. Sloan foundation.


\begin{thebibliography}{10}



\bibitem{Su80} W. P. Su {\em et al.},
Phys. Rev. B{\bf 22}, 2099 (1980).
\bibitem{Tsui82}D. C. Tsui {\em et al.},
Phys. Rev. Lett. {\bf 48}, 1559 (1982).
\bibitem{Laughlin83}R. B. Laughlin, Phys. Rev. Lett. {\bf 50}, 1395 (1983).
\bibitem{Jackiw76}
R. Jackiw and C. Rebbi, Phys. Rev. D {\bf 13}, 3398 (1976).



\bibitem{Kitaev03}A. Kitaev, Ann. of. Phys. {\bf 303}, 2(2003);
{\bf 321}, 2 (2006).


\bibitem{Duan03}L. M. Duan {\em et al.},
Phys. Rev. Lett. {\bf 91}, 090402 (2003).

\bibitem{Buchler05}H. P. Buchler {\em et al.}, Phys. Rev. Lett. {\bf 95}, 040402 (2005).

\bibitem{Micheli06} A. Micheli {\em et al.}, Nature Phys. {\bf 2}, 341 (2006).

\bibitem{Demler02} E. Demler, F. Zhou, Phys. Rev. Lett. {\bf 88}, 163001
(2002).

\bibitem{Mukerjee06} S. Mukerjee {\em et al.}, Phys. Rev. Lett. {\bf 97},
120406 (2006).

\bibitem{Semenoff07}
Gordon Semenoff and Fei Zhou,
Phys. Rev. Lett. {\bf 98}, 100401 (2007).

\bibitem{Thouless98}
For general discussions, also see D. J. Thouless, {\it Topological
Quantum Numbers in Nonrelativistic Physics}(World Scentific, 1998).


\bibitem{Hadzibabic06} Z. Hadzibabic {\em et al.}, Nature {\bf
441}, 1118 (2006).


\bibitem{Wen04}X. G. Wen, {\it Quantum Theory of
Many-body Systems}(Oxford University Press, 2004).


\bibitem{Stenger98} J. Stenger {\em et al.}, Nature (London){\bf 396}, 345 (1998).


\bibitem{Zhou03}
F. Zhou and M. Snoek, Ann. Phys. {\bf 308}, 692 (2003);
M. Snoek and F. Zhou, Phys. Rev. B {\bf 69}, 094410 (2004).



\bibitem{Zhou01}
F. Zhou, Phys. Rev. Lett. {\bf 87}, 080401(2001).


\bibitem{Romans04} M. W. J. Romans, Phys. Rev. Lett. {\bf 93},
020405 (2004).



\bibitem{Song08} J. L. Song, J. Zhang and F. Zhou, unpublished.

\bibitem{Fradkin79} E. Fradkin and S. Shenker, Phys. Rev. {\bf D 19},
 3682(1979).
\bibitem{Toner93} P. E. Lammert {\em et al.}, Phys. Rev.
 Lett. {\bf 70}, 1650(1993).


\bibitem{Wilson74} K. G. Wilson, Phys. Rev. D {\bf 10}, 2445(1974).


\bibitem{Madison00} K. W. Madison {\em et al.},
Phys. Rev. Lett. \textbf{84}, 806 (2000); K. W.
Madison {\em et al.}, Phys. Rev. Lett. \textbf{86}, 4443 (2001).

\bibitem{Haljan01} P. C. Haljan {\em et al.}, Phys. Rev.
Lett. {\bf 87}, 210403 (2001).

\bibitem{AboShaeer01} J. R. Abo-Shaeer {\em et al.}, Science
{\bf 292}, 479 (2001).

\bibitem{Fetter01} A. L. Fetter and A. A. Svidzinsky,
J. Phys.: Condens. Matt. {\bf 13}, R135 (2001).

\bibitem{Dalfovo01}F. Dalfovo, S. Stringari, Phys. Rev. A {\bf 63},
011601 (2000).


\bibitem{Tsubota02} M. Tsubota {\em et al.}, Phys. Rev. A
\textbf{65}, 023603 (2002).

\bibitem{Isoshima02}
T. Isoshima, K. Machida, Phys. Rev. A \textbf{66}, 023602(2002).




\bibitem{Feder00} D. Feder {\em et al.}, Phys. Rev. A{\bf 61}, 011601 (2000).
The hydrodynamic breakdown length $L_{B}$ is about
$(L^4_{h.o.}/2R_{TF})^{1/3}$, which in our case turns out to be about $2a$
($a$ is the lattice constant.).


\bibitem{Simula02} T. P. Simula {\em et al.}, Phys. Rev. A{\bf 66}, 035601
(2002); T. Mizushima {\em et al.}, Phys. Rev. A {\bf 64}, 043610 (2001).






\bibitem{Andrew08}Anchun Ji {\em et al.}, Phys. Rev. Lett. {\bf 101},
010402(2008).





\bibitem{Chevy00}
F. Zambelli and S. Stringari, Phys. Rev. Lett. {\bf 81}, 1754 (1998);
F. Chevy {\em et al.}, Phys. Rev. Lett.{\bf 85}, 2223 (2000).
See also discussions on liquid helium in W. F. Vinen, Nature (London) {\bf 181}, 1524
(1958).

\end{thebibliography}
\end{document}